\documentclass[a4paper]{ring}
\usepackage{graphicx}
\usepackage{natbib}

\newcommand{\mdot}{\dot{m}}

\begin{document}

\title{A multi-flow model for microquasars}

\author{Pierre-Olivier Petrucci\inst{1} ,Jonathan Ferreira\inst{1},
  Gilles Henri\inst{1}, Ludovic  Saug\'e\inst{1,2} \and Guy
  Pelletier\inst{1}}

\institute{Laboratoire d'Astrophysique, Observatoire de Grenoble BP53,
  F-38041 Grenoble cedex 9, France \and Institut de Physique Nucleaire de
  Lyon, 43 bd 11 novembre 1918, F-69622 Villeurbanne cedex, France}

\titlerunning{A multi-flow model for microquasars}
\authorrunning{Petrucci et al.}

\maketitle

\centerline{\bf Abstract}
{We present a new picture for the central regions of Black Hole X-ray
Binaries.  In our view, these central regions have a multi-flow
configuration which consists in (1) an outer standard accretion disc down
to a transition radius $r_J$, (2) an inner magnetized accretion disc
below $r_J$ driving (3) a non relativistic self-collimated
electron-proton jet surrounding, when adequate conditions for pair
creation are met, (4) a ultra relativistic electron-positron beam.

This accretion-ejection paradigm provides a simple explanation to the
canonical spectral states, from radio to X/$\gamma$-rays, by varying the
transition radius $r_J$ and disc accretion rate $\dot m$
independently. Large values of $r_J$ and low $\dot m$ correspond to
Quiescent and Hard states. These states are characterized by the presence
of a steady electron-proton MHD jet emitted by the disc below $r_J$. The
hard X-ray component is expect to form at the jet basis. When $r_J$
becomes smaller than the marginally stable orbit $r_i$, the whole disc
resembles a standard accretion disc with no jet, characteristic of the
Soft state. Intermediate states correspond to situations where $r_J \ga
r_i$. At large $\dot m$, an unsteady pair cascade process is triggered
within the jet axis, giving birth to flares and ejection of relativistic
pair blobs. This would correspond to the luminous intermediate state,
sometimes referred to as the Very High state, with its associated
superluminal motions. Some features such as possible hysteresis and the
presence of quasi-periodic oscillations could be also described in this
paradigm.

}

\section{A novel framework for BH XrBs}

\subsection{General picture}
\label{mainpict}
\begin{figure*}
\includegraphics[width=\textwidth]{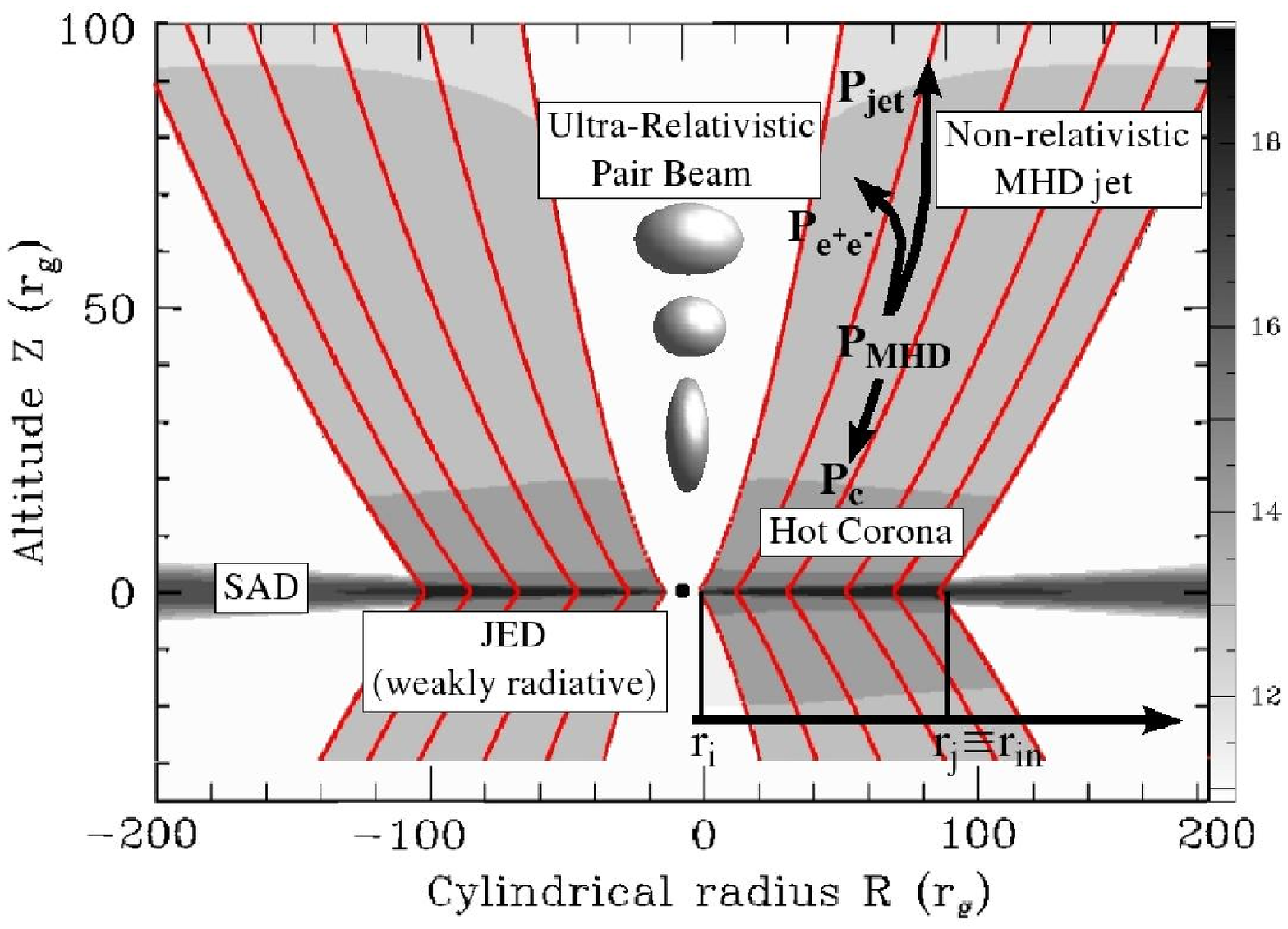}
\caption{A Standard Accretion Disc (SAD) is established down to a radius
$r_J$ which marks the transition towards a low radiative Jet Emitting
Disc (JED), settled down to the last stable orbit. The JED is driving a
mildly relativistic, self-collimated electron-proton jet which, when
suitable conditions are met, is confining an inner ultra-relativistic
electron-positron beam. The MHD power $P_{MHD}$ flowing from the JED acts
as a reservoir for (1) heating the jet basis (radiating as a moving
thermal corona with power $P_c$), (2) heating the inner pair beam
($P_{e^+e^-}$) and (3) driving the compact jet ($P_{jet}$). Field lines
are drawn in black solid lines and the number density is shown in
greyscale ($\log_{10} n/\mbox{m}^{-3}$). This magnetic accretion-ejection
structure solution was computed with $\xi=0.01$, $\varepsilon=0.01$ and
with $m=10$ and $\dot m((r_J)=0.01$ (see text).}
\label{fig:smae2flow}
\end{figure*}

We assume that the central regions of BH XrB are composed of four
distinct flows: two discs, one outer "standard" accretion disc (hereafter
SAD) and one inner jet emitting disc (hereafter JED), and two jets, a
non-relativistic, self-confined electron-proton MHD jet and, when
adequate conditions for pair creation are met, a ultra-relativistic
electron-positron beam. A sketch of our model is shown in
Fig.~\ref{fig:smae2flow} while the four dynamical components are
discussed separately below. This is an extended version of the "two-flow"
model early proposed for AGN and quasars
\citep{pel88b,sol89,pel89,hen91,pel92} to explain the highly relativistic
phenomena such as superluminal motions observed in these sources. This
model provides a promising framework to explain the canonical spectral
states of BH XrBs mainly by varying the transition radius $r_J$ between
the SAD and the JED. This statement is not new and has already been
proposed in the past by different authors
(e.g. \citealt{esi97,bel97,liv03,kin04}) but our model distinguishes
itself from the others by the consistency of its disc--jet structure and
by the introduction of a new physical component, the ultra-relativistic
electron-positron beam, that appears during strong outbursts.

We believe that jets from BH XrBs are self-collimated because they follow
the same accretion-ejection correlation as in AGN
\citep{cor03,fen03,mer03}. This therefore implies the presence of a large
scale vertical field anchored somewhere in the accretion disc (the JED)
and we assume that this large scale $B_z$ has the same polarity. The
presence of a large scale vertical field threading the disc is however
not sufficient to drive super-Alfv\'enic jets. This field must be close
to equipartition as shown by \cite{fer95} and \cite{fer97}. The reason is
twofold. On one hand, the magnetic field is vertically pinching the
accretion disc so that a (quasi) vertical equilibrium is obtained only
thanks to the gas and radiation pressure support. As a consequence, the
field cannot be too strong. But on the other hand, the field must be
strong enough to accelerate efficiently the plasma right at the disc
surface (so that the slow-magnetosonic point is crossed smoothly). These
two constraints can only be met with fields close to equipartition.\\

An important local parameter is therefore the disc magnetization $\mu =
B_z^2/(\mu_o P_{tot})$ where $P_{tot}$ includes the plasma and radiation
pressures. In our picture, a SAD is established down to a radius $r_J$
where $\mu$ becomes of order unity. Inside this radius, a JED with $\mu
\sim 1$ is settled.  At any given time, the exact value of $r_J$ depends
on highly non-linear processes such as the interplay between the amount
of new large scale magnetic field carried in by the accreting plasma
(eg. coming from the secondary) and turbulent magnetic diffusivity
redistributing the magnetic flux already present. These processes are far
to be understood. For the sake of simplicity, we will treat in the
following $r_J$ as a free parameter that may vary with time (see
Section~3).

\subsection{The outer SAD}
\label{SAD}

Accretion requires the presence of a negative torque extracting angular
momentum. In a SAD this torque is assumed to be of turbulent origin and
provides an outward transport of angular momentum in the radial
direction. It has been modeled as an "anomalous" viscous torque of
amplitude $\sim - \alpha_v P_{tot}/r$, where $\alpha_v$ is a small
parameter \citep{sha73}. The origin of this turbulence is now commonly
believed to arise from the magneto-rotational instability or MRI
\citep{bal91}. The MRI requires the presence of a weak magnetic field
($\mu < 1$) and is quenched when the field is close to equipartition. We
make the conjecture that a SAD no longer exists once $\mu$ reaches
unity. Indeed, it can be easily shown that one may reasonably expect
$\mu$ to increase towards the center \citep{fer05}. Whenever a BH XrB
reaches $\mu \simeq 1$ at a radius $r_J > r_i$, $r_i$ being the last
marginally stable orbit, the accretion flow changes its nature to a JED.
To summarize, the accretion flow at $r> r_J$ is a SAD with $\mu \ll 1$
fueled by the companion's mass flux and driving no outflow (constant
accretion rate $\dot M_a$).

\subsection{The inner JED}
\label{JED}

This inner region with $\mu \sim 1$ is fueled by the SAD at a rate $\dot
M_{a,J} = \dot M_a (r_J)$. Since it undergoes mass loss, {we parametrize}
the JED accretion rate { following}:
\begin{equation}
\dot M_a(r) = \dot M_{a,J} \left (\frac{r}{r_J} \right )^\xi
\end{equation}
where $\xi$ measures the local ejection efficiency \citep{fer93a}. The
global energy budget in the JED is $P_{acc,JED} = 2 P_{rad,JED} + 2
P_{MHD}$ where $P_{MHD}$ is the MHD Poynting flux feeding a jet, whereas
the liberated accretion power writes
\begin{equation}
\label{paccjed}
P_{acc,JED} \simeq \frac{GM \dot M_{a,J}}{2r_i} \left [ \left (
\frac{r_i}{r_J} \right )^\xi - \frac{r_i}{r_J} \right ]
\label{eq:bilan}
\end{equation}
The dynamical properties of a JED have been extensively studied in a
series of papers (see \citealt{fer02} and references therein). The ratio
at the disc midplane of the jet torque to the turbulent "viscous" torque
is
\begin{equation}
\Lambda \sim \frac{ B_\phi^+ B_z/\mu_o h}{\alpha_v P_{tot}/r} \sim \frac{B_\phi^+ B_z}{\mu_o P_{tot}} \frac{r}{\alpha_v h}
\end{equation}
It is straightforward to see that the necessary condition to drive jets
(fields close to equipartition) from Keplerian discs leads to a dominant
jet torque. In fact, it has been shown that steady-state ejection
requires $\Lambda \sim r/h \gg 1$ \citep{fer97,cas00a}.

This dynamical property has a tremendous implication on the JED
emissivity since it can be shown that the total luminosity $2P_{rad,JED}$
of the JED is only a fraction $1/(1+\Lambda)$ of the accretion disc
liberated power $P_{acc,JED}$ \citep{fer05}. In consequence, the JED is
weakly dissipative while powerful jets are being produced regardless of
the nature of the central object. As a consequence, the flux emitted by
the JED is expected to be unobservable with respect to that of the outer
SAD.  Thus, the values of the "disc inner radius" ($r_{in}$) and "disc
accretion rate" observationally determined from spectral fits must be
understood here as values at the transition radius, namely $r_{in} \equiv
r_J$ and $\dot m \equiv \dot m(r_J)$: the optically thick JED is
spectrally hardly visible.

\subsection{Non-relativistic electron-proton jets from JEDs}
\label{MAES}

The ejection to accretion rate ratio in a JED writes $2 \dot M_{jet}/\dot
M_{a,J} \simeq \xi \ln (r_J/r_i)$. In principle, the ejection efficiency
$\xi$ can be observationally deduced from the terminal jet speed. Indeed,
the maximum velocity reachable along a magnetic surface anchored on a
radius $r_o$ (between $r_i$ and $r_J$) is $u_\infty \simeq \xi^{-1/2}
\sqrt{GM/r_o}$ in the non-relativistic limit (see \citealt{fer97} for
relativistic estimates). Although a large power is provided to the
ejected mass (mainly electrons and protons), the mass loss ($\xi$) is
never low enough to allow for speeds significantly relativistic required
by superluminal motions: MHD jets from accretion discs are basically non
or only mildly relativistic with $u_\infty \sim 0.1 - 0.8\ c$
\citep{fer97}. This is basically the reason why they can be efficiently
self-confined by the magnetic hoop stress. Indeed, in relativistic flows
the electric field grows so much that it counteracts the confining effect
due to the toroidal field. This dramatically reduces the self-collimation
property of jets \citep{bog01,bog01b,pel04}.\\

In our framework, jets from magnetic accretion-ejection structure
(hereafter MAES) have two distinct spectral components detailed below:

\subsubsection{A non-thermal extended jet emission}

We expect a small fraction of the jet power $P_{jet}$ to be converted
into particles, through first and/or second order Fermi acceleration,
populating the MHD jet with supra-thermal particles. These particles are
responsible for the bulk emission of the MHD jet. This is similar to
models of jet emission already proposed in the literature
\citep{fal95a,vad01,mark01,mark03,mark04b,fal04}.  In these models, the
jet is assumed to be radiating self-absorbed synchrotron emission in the
radio band (producing a flat or even inverted radio spectrum) becoming
then optically thin in the IR-Optical bands and providing a contribution
up to the X/$\gamma$-rays. 

\subsubsection{A thermal jet basis}

Jet production relies on a large scale magnetic field anchored on the
disc as much as on MHD turbulence triggered (and sustained) within
it. This implies that small scale magnetic fields are sheared by the disc
differential rotation, leading to violent release of magnetic energy at
the disc surface and related turbulent heat fluxes (e.g. \citealt{gal79,
hey89a,sto96, mer02}). The energy released is actually tapping the MHD
Poynting flux flowing from the disc surface. We can safely assume that a
fraction $f$ of it would be deposited at the jet basis, with a total
power $P_c = f P_{MHD}$. The dominant cooling term in this optically thin
medium is probably comptonization of soft photons emitted by the outer
SAD (with a small contribution from the underlying JED). These are
circumstances allowing a thermal plasma to reach a temperature as high as
$\sim 100$ keV, \citep{kro95,mah97,esi97}. This plasma being at the base
of the jet, it will have a vertical proper motion. Then its spectral
behavior is expected to be close to that of a dynamic corona
\citep{mal01}.\\

\subsection{The inner ultra-relativistic pair beam}
\label{pairbeam}

Since the large scale magnetic field driving the self-confined jet is
anchored onto the accretion disc which has a non zero inner radius, there
is a natural hole on the axis above the central object with no baryonic
outflow (this also holds for neutron stars). This hole provides a place
for pair production and acceleration with the outer MHD jet acting as a
sheath that confines and heats the pair plasma. This is the microquasar
version of the "two flow" model that has been successfully applied to the
high energy emission of relativistic jets in AGNs
\citep{hen91,mar95,mar98,ren98}. \\

The $e^+-e^-$ plasma is produced by $\gamma-\gamma$ interaction, the
$\gamma$-ray photons being initially produced by a few relativistic
particles by Inverse Compton process, either on synchrotron photons
(Synchrotron Self Compton or SSC) or on disc photons (External Inverse
Compton or EIC).

It is well known that above 0.5 MeV photons can annihilate with
themselves to produce an electron-positron pair. Usually, pairs are
assumed to cool once they are formed, producing at turn non thermal
radiation. Some of this radiation can be absorbed to produce new pairs,
but the overall pair yield never exceeds 10 \%.  A key point of the
two-flow model however is that the MHD jet launched from the disc can
carry a fair amount of turbulent energy, most probably through its MHD
turbulent waves spectrum. A fraction of this power can be transferred to
the pairs ($P_{e^+e^-} << P_{MHD}$). Thus the freshly created pairs can
be continuously reheated, triggering an efficient pair runaway process,
leading to a dense pair plasma \citep{hen91}.\\

As we said, reacceleration is balanced by cooling through the combination
of synchrotron, SSC and EIC processes. Synchrotron and SSC emission are
quasi isotropic in the pair frame, but the external photon field is
strongly anisotropic. The pair plasma will then experience a strong bulk
acceleration due to the recoil term of EIC, an effect also known as the
"Compton Rocket" effect \citep{ode81,ren98}. As shown in previous works,
this "rocket" effect is the key process to explain relativistic motion
\citep{mar95, ren98}. For example, values of 5 to 10 can be easily
reached in near-Eddington accretion regime around stellar black holes
\citep{ren98}.\\

Producing this pair plasma requires thus altogether a strong MHD jet, a
radiative non-thermal component extending above the MeV range and a
minimal $\gamma-\gamma$ optical depth, namely $\tau_{\gamma \gamma} \sim
1$.  Using simple estimates, the optical depth $\tau_{\gamma \gamma}$ for
absorbing photons with energy $E_\gamma= \varepsilon m_{\rm e} c^2$ is
approximately, for a spherical source of radius $R$ filled by soft
photons with a power-law distribution $\nu L_{\nu} = E L_E = L_0
(E/E_0)^{-\Gamma+2}$ (where $\Gamma$ is the soft photon index):
\begin{eqnarray}
& &\tau_{\gamma \gamma} (E_\gamma) = 0.7\times 262^{(2.5-\Gamma)} \left(
\frac{L_0}{0.1 L_{Edd}} \right)\nonumber\\ && \times\left(\frac{R}{30
r_g}\right)^{-1} \left(\frac{E_0}{1 \rm{ keV}}\right)^{\Gamma-2}
\left(\frac{E_\gamma}{1 \rm{ MeV}}\right)^{\Gamma-1} 
\label{eqtaugg}
\end{eqnarray}
where $r_g= GM/c^2$. Thus, good conditions for pair creation require high
luminosity, small size and high energy ($>$ MeV) photons.\\
 
It is noteworthy that the pair beam is intrinsically highly variable and
subject to an intermittent behavior. Indeed, once the pair creation is
triggered, a regulation mechanism must occur to avoid infinite power of
the pair plasma and limit the pair run-away. This is probably realized by
the quenching of the turbulence ($P_{e^+e^-}$ vanishes) when most of its
energy is suddenly tapped by the catastrophic number of newly created
pairs. These pairs will therefore simply expand freely, confined by the
heavier MHD jet. One would then expect a flare in the compact region,
followed by the ejection of a superluminal radio component, analoguous to
those observed in AGNs (Saug\'e \& Henri 2005, A\&A submitted). Such a
situation can repeat itself as long as the required physical conditions
are met. Alternatively, it may also that the formation of a dense pair
beam destroys the surrounding MHD jet, explaining the disappearance of
the compact jet after a strong ejection event.


\section{Canonical spectral states of X-ray binaries}
\label{specstate}

\subsection{The crucial roles of $r_J$ and $\mdot$}

From Section~1, it is clear that the spectral appearance of a BH XrB
critically depends on the size of the JED relative to the SAD, namely
$r_J$. As stated before, $r_J$ is the radius where the disc magnetization
$\mu = B_z^2/(\mu_o P_{tot})$ becomes of order unity.  Thus, $r_J$
depends on two quantities $P_{tot}(r,t)$ and $B_z(r,t)$. The total
pressure is directly proportional to $\mdot$ since $P_{tot}= \rho
\Omega_k^2 h^2 \propto \mdot m^{-1}r^{-5/2}$. As a consequence, any
variation of the accretion rate in the outer SAD implies also a change in
the amplitude of the total pressure. But we have to assume something
about the time evolution of the large scale magnetic field threading the
disc. The processes governing the amplitude and time scales of these
adjustments of $r_J$ to a change in $\mdot$ are far too complex to be
addressed here. They depend on the nature of the magnetic diffusivity
within the disc but also on the radial distribution of the vertical
magnetic field. We will simply assume in the following that $r_J$ and
$\dot m$ are two independent parameters. In that respect, our view is
very different from that of \cite{esi97,mah97} who considered only the
dependency of $\dot m$ to explain the different spectral states of BH
XrBs.

\begin{figure*}
\begin{center}
\includegraphics[height=18cm]{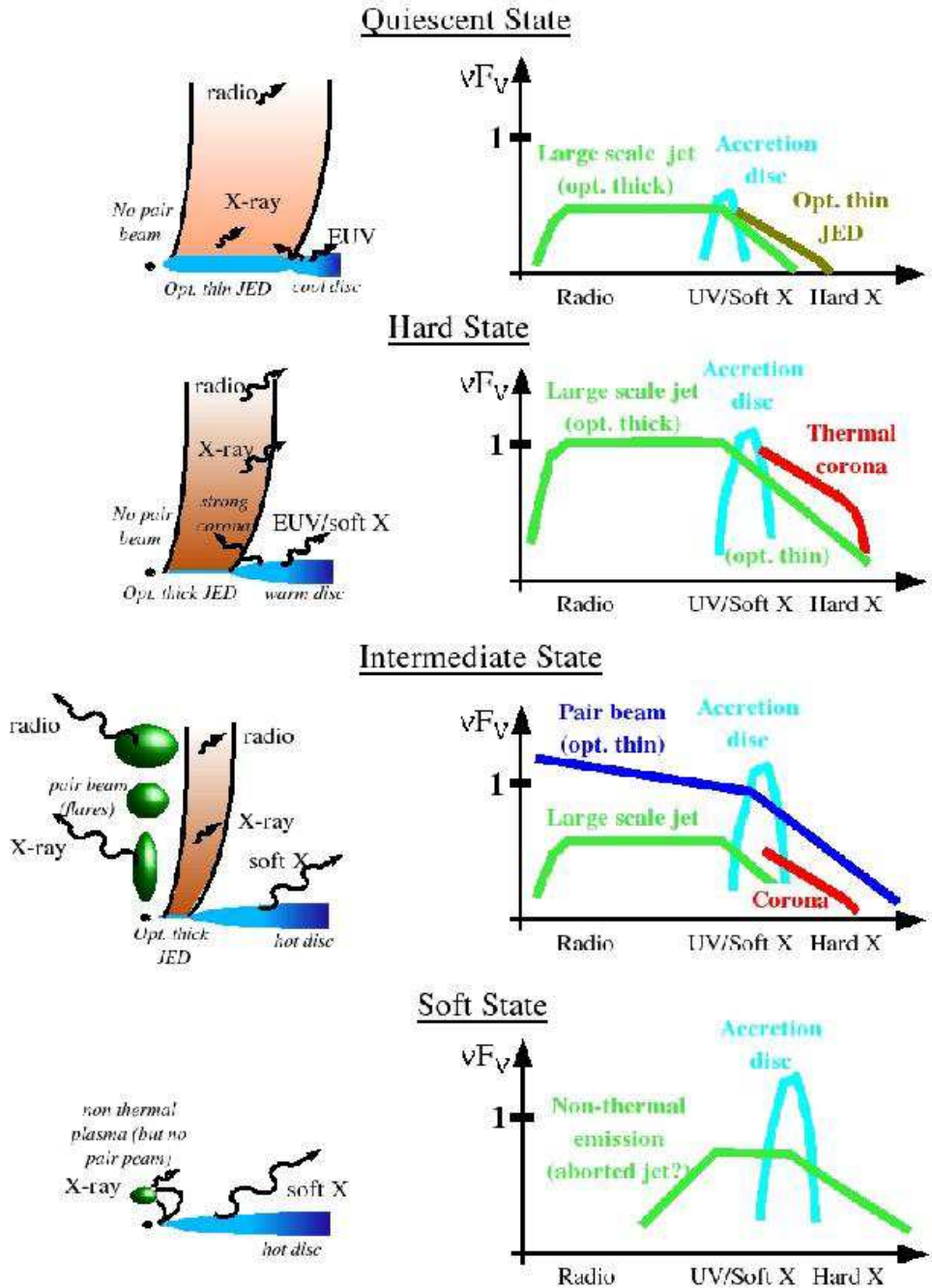}
\end{center}
\caption{The canonical spectral states of X-ray binaries
  (cf. Sect. \ref{specstate} for more details). {\bf (a)} Quiescent state
  obtained with a low $\mdot$ and a large $r_J$: the Jet Emitting Disc
  (JED) occupies a large zone in the accretion disc. {\bf (b)} Hard state
  with much larger $\mdot$ and smaller $r_J$: the pair creation threshold
  is still not reached. {\bf (c)} Luminous Intermediate state between the
  Hard and the Soft states: the high disc luminosity combined with the
  presence of a MHD jet allows pair creation and acceleration along the
  axis, giving birth to flares and superluminal ejection events.{\bf (d)}
  Soft state when there is no zone anymore within the disc where an
  equipartition field is present: no JED, hence neither MHD jet nor pair
  beam.}
\label{diffstate}
\end{figure*}

\subsection{The Quiescent state} 

This state is characterized by a very low accretion rate ($\mdot$ as low
as $\sim$ 10$^{-9}$) with a hard X-ray component. The ADAF model has been
successfully applied to some systems with a large transition radius
between the ADAF and the outer standard disc, namely $r_{tr} \sim
10^3-10^4\ r_g$ (e.g. \citealt{nar96,ham97}). However, such a model does
not account for jets and their radio emission, even though XrBs in
quiescence seem also to follow the radio/X-ray correlation
(e.g. \citealt{fen03,gal04,gal05}).

Within our framework, a BH XrB in quiescence has a large $r_J$, so that a
large zone in the whole disc is driving jets (Fig.~\ref{diffstate}a). The
low $\dot m$ provides a low synchrotron jet luminosity, while the JED is
optically thin, producing a SED probably very similar to that of an
ADAF. We thus expect $r_J \sim r_{tr}$. The weak MHD Poynting flux
prevents the ignition of the pair cascade process and no pair beam is
produced.

\subsection{The Hard state} 

Within our framework, the JED is now more limited radially than in the
Quiescent state, namely $r_J \sim 40-100\ r_g$
(Fig.~\ref{diffstate}b). This transition radius corresponds to the inner
disc radius $r_{in}$ as obtained within the SAD framework \citep{zdz04b}.
The low velocity of the plasma expected at the jet basis is in good
agreement with recent studies of XrBs in Hard state
\citep{mac03b,gal03}. It can also explain the apparent weakness of the
Compton reflection \citep{zdz99,gil99} as already suggested by \citet[see
also \citealt{bel99,mal01}]{mark03} and tested by \cite{mark04}. In any
case, the JED intrinsic emission is weak with respect to that of the
outer standard disc: most of the accretion power flows out of the JED as
an MHD Poynting flux. Nevertheless, the threshold for pair creation is
still not reached and there is no pair beam, hence no superluminal
motion. The MHD power is therefore shared between the jet basis, whose
temperature increases (the thermal "corona") producing X-rays, and the
large-scale jet seen as the persistent (synchrotron) radio emission.

\subsection{The Soft state} 

Our interpretation of the Soft state relies on the disappearance of the
JED, i.e. when $r_J$ becomes smaller or equal to $r_i$
(Fig.~\ref{diffstate}c). Depending on the importance of the magnetic flux
in the disc, this may occur at different accretion rates. Thus, the
threshold in $\dot m$ where there is no region anymore in the disc with
equipartition fields may vary. The whole disc adopts therefore a radial
structure akin to the standard disc model. Since no MHD jet is produced,
all associated spectral signatures disappear. Even if pair production may
take place (when $\dot m$ is large), the absence of the confining MHD jet
forbids the pairs to get warm enough and be accelerated: no superluminal
motion should be detected.

Note also that the presence of magnetic fields may be the cause of
particle acceleration responsible for the weak hard-energy tail (McCR03,
\citealt{zdz04} and references therein).

\subsection{Intermediate states}
 
This state has been first identified at large luminosities ($L> 0.2\
L_{Edd}$) and was initially called Very High state. However, high
luminosity appeared to not be a generic feature since it has be observed
at luminosities as low as $0.02\ L_{Edd}$ (McCR03,
\citealt{zdz04}). Therefore, the most prominent feature is that these
states are generally observed during transitions between Hard and Soft
states. Within our framework, they correspond to geometrical situations
where $r_J$ is small but remains larger than $r_i$
(Fig.~\ref{diffstate}d). The flux of the outer standard disc is thus
still important while the JED is disappearing. The consequences on the
spectral shape are not straigtforward since the importance of the
different spectral components relative to each other depends on the
precise values of $r_J$ and $\dot m$. Such study is out of the scope of
the present paper and will be detailed elsewhere. \\

The crucial point however is that, in our framework, luminous
intermediate states (the so-called Very High State or VHS) with high
$\dot m$ provide the best conditions for the formation of the
ultra-relativistic pair beam, as described in details in
Sect.~\ref{pairbeam}: (1) a high luminosity, (2) a high energy steep
power law spectrum extended up to the $\gamma$-ray bands and (3) the
presence of the MHD jet . The two first characteristics enable a
$\gamma-\gamma$ opacity larger than unity (cf. Eq.~\ref{eqtaugg} of
Sect.~\ref{pairbeam}), while the MHD jet allows to confine the pair beam
and maintain the pairs warm, a necessary condition to trigger a pair
runaway process. The total emission would be then dominated by the
explosive behavior of the pairs, with the sudden release of blobs. Each
blob produced in the beam first radiates in X and $\gamma$-ray,
explaining the hard tail present in this state, and then, after a rapid
expansion, produces the optically thin radio emission. The production of
a series of blobs can even result in an apparently continuous spectrum,
from radio to X/$\gamma$-rays. This pair beam would also explain the
superluminal ejections observed during this state in different objects
(e.g. \citealt{sob00,han01}). We conjecture that the exact moment where
this occurs corresponds to the crossing of the "jet line" recently
proposed by \cite{fen04c} (see also \citealt{cor04}). This corresponds to
a transition from the "hard" intermediate state to the "soft" one.\\

\section{Time evolution of BH XrBs}
\label{evolution}
\begin{figure*}
\centering
\includegraphics[width=0.7\textwidth]{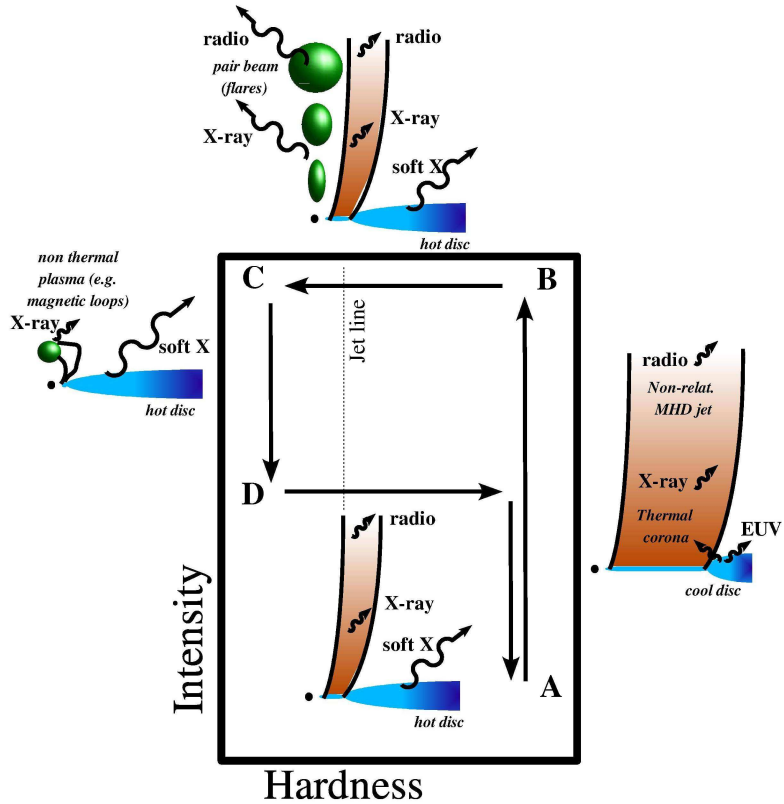}
\caption{Schematic Hardness--Intensity diagram as it is generally
  observed in XrBs in outbursts (this figure is clearly inspired by
  Fig. 7 of Fender et al. 2004). During such outbursts, the objects
  follow the A-B-C-D sequence before turning back to A at the end of the
  outburst. The interpretation of this diagram within our framework is
  detailed in Sect. \ref{evolution}. 
}
\label{fig:hid}
\end{figure*}

The evolution with time of a BH XrB has been reported in
Fig.~\ref{fig:hid} (Ferreira et al. in preparation). This is a synthetic
Hardness--Intensity diagram (hereafter HID) as it is generally observed
in XrBs {in outbursts} (e.g. Belloni et al. 2005; Fender et al. 2002,
2004).  { During such outbursts, the objects follow the A-B-C-D sequence
before turning back to A at the end of the outburst}.  We detailed below
the interpretation of this diagram in the framework of our model. We have
also overplotted on Fig. \ref{fig:hid} the different sketches of our
model at different phases (this figure is clearly inspired by Fig. 7 of
Fender et al. 2004).\\

\subsection{Ascending the Right Branch: }

Let us start at a Low/Hard State located at the bottom of the HID right
branch (in A in Fig. \ref{fig:hid}). In our view, such state would
correspond to a JED extending up to typically $r_J \sim 10^2 r_g$.  This
considerably lowers the emission from the inner radii of the SAD
producing a UV/soft X-ray excess. The hard (1-20 keV) power-law component
of photon index $\Gamma \sim 1.7$ is attributed to the warm thermal
plasma at the base of the jet. The non relativistic MHD jet then produces
the persistent IR and radio synchrotron emission.

\subsection{The Top Horizontal Branch}
\paragraph{\bf Before the jet line:}
{{Arriving in B} we assume that $r_J$ starts decreasing rapidly. Then,
the MAES undergoes an outside-in transition to a SAD. The BH XrBs enter
the high intermediate state.\\

The flux of the outer standard disc then increases while the JED is
decreasing. Under such circumstances, the MHD Poynting flux released by
the JED is still important (through the large $\mdot$ that characterizes
this part of the HID) but the MHD jet itself fills a smaller volume, a
direct consequence being a weaker emission of the thermal "corona" and
the non-thermal MHD jet emission with respect to what it is while in the
Hard state.\\

\paragraph{\bf At the jet line:}

During its evolution along this top horizontal branch the system can
reach a critical phase where the conditions for a strong pair production,
inside the MHD jet structure, are fulfilled.

In this case, we expect an explosive behaviour of the pairs, with the
sudden release of blobs. The emission of these blobs, first in X and
$\gamma$-ray and then, after a rapid expansion, in IR and radio, will
probably dominate the broad band spectrum, producing the hard X-ray tail
and the optically thin radio emission present in this state. The
production of a series of blobs can even result in an apparently
continuous spectrum, from radio to X/$\gamma$-rays.\\ Remarkably, there
is no evidence of steady radio jets during this phase but it is generally
associated with radio and X-ray flares and/or superluminal sporadic
ejections (e.g. \citealt{sob00,han01}).\\

We note that the rapid increase of the pair beam pressure in the inner
region of the MHD jet, during e.g. a strong outburst, may dramatically
perturb the MHD jet production. Indeed, a huge pair pressure at the axis
may enforce the magnetic surfaces to open dramatically, thereby creating
a magnetic compression on the disc (actually the JED) so that no more
ejection is feasible. Alternatively, it is also possible that the racing
of the pair process completely wears out the MHD Poynting flux released
by the JED, suppressing the jet emission or even the jet itself. Whatever
occurs (i.e. jet destruction or jet fading), we expect a suppression of
the steady jet emission when a large outburst sets in. Interestingly, the
detailed spectral and timing study of the radio/X-ray emission of four
different black hole binaries during a major radio outburst
\citep{fen04c} shows a weakening and softening of the X-ray emission
as well as a the quenching of the radio emission after the burst. This is
in good agreement with our expectations since the cooling of the pair
beam should indeed results in a flux decrease and a softening of its
spectrum.

It is also worth noting that within our picture, we do not expect all BH
XrBs to reach a ``pair production'' phase along the top branch. Indeed,
it requires a rather high accretion rate so that the gamma-gamma opacity
reaches unity in order to trigger the pair cascade process. Moreover, a
large accretion rate implies a large disc pressure. Since the pair beam
requires to be surrounded by a MHD jet (for confinment and energetic
reasons), this also implies a large magnetic flux in the disc able to
match this increase in $P_{tot}$. In practice, one needs to have $r_J
\sim 10 r_g$ at these high levels of accretion activity
(cf. Eq. \ref{eqtaugg}).

\paragraph{\bf After the jet line:}

We assume that $r_J$ is still decreasing. We therefore expect the total
disappearance of the JED and its MHD jets when $r_J \rightarrow r_i$,
thereby also causing the end of the pair beam { (if present)}. The inner
regions of the BHXB are a SAD with probably a magnetically active
"corona". Indeed, it must be noted that the situation might be slightly
more complex than a mere SAD because of the presence of a concentrated
magnetic flux. No steady MHD ejection can be produced from the SAD but
unsteady events could always be triggered. This is maybe the reason why
this region in the HID seems to harbor complex variability phenomena
\citep{bel05,nesp03}.


\subsection{Descending the Left Branch}

{ When XrBs reach the left vertical branch (point C in
Fig. \ref{fig:hid}), $r_J$ is smaller than the inner disc radius i.e. the
JED and the MHD jet have completely disappeared.  The whole disc adopts
therefore a radial structure akin to the standard disc model and we enter
into the so-called soft state (also called thermal dominant state McCR03)
where the spectra are dominated by strong disc emission. The descent from
C to D correspond to a decrease in intensity i.e. by a decrease of the
accretion rate. This is the beginning of the fading phase of the
outburst. In our framework $r_J$ keeps smaller than $r_i$.

We note also that we still expect the presence of magnetic fields that
may be the cause of particle acceleration responsible for the weak
hard-energy tail generally observed in this state (McCR03, ZG04 and
references therein).\\}






\subsection{The Low Horizontal Branch}

In D $r_J$ begins to increase again. Thus, according to this conjecture,
there is an inside-out build up of a JED. Self-collimated electron-proton
jets could be produced right away. This means an increase of $r_j$, the
reappearance of the non-thermal MHD jet and the thermal corona at its
basis and a decrease of the SAD emission. But, contrary to the Top
Horizontal Branch, the accretion rate is now too low to allow the
production of a pair beam. { Concequently we do not expect superluminal
motions during this phase.}


\subsection{The Quiescent State}
When $r_J$ reaches the same value as in the Low/Hard State the system is
ready for another duty cycle. But much larger values of $r_J$ can be
obtained, following the same process, if the accretion rate undergoes a
strong decline towards quiescent levels. 

In our picture, the Quiescent State should be described by a large radial
extension covered by an underluminous JED with $\mu \sim 1$.
%
A large zone in the whole disc is therefore driving jets as inferred from
some observations (e.g. \citealt{fen03,gal04}). Jets are radiating
through synchrotron emission and produce an optically thick radio
spectrum (although very low). On the other hand, for such a low accretion
rate, the JED is expected to be in large part optically thin. The
computation of the SEDs have been undergone (Petrucci et al. in
preparation).

%

\section{Summary and concluding remarks}

We present in this paper a new paradigm for the accretion-ejection
properties of Galactic Black Hole X-ray binaries. We assume the existence
of a large scale magnetic field of bipolar topology in the innermost disc
regions. Such a field allows for several dynamical phenomena to occur
whose relative importance determine the observed spectral state of the
binary. The dynamical constituents are: (1) an outer standard accretion
disc (SAD) for $r> r_J$, (2) an inner Jet Emitting Disc (JED) for $r<r_J$
driving (3) a self-collimated non-relativistic electron-proton
surrounding, when adequate conditions are met, (4) a ultra-relativistic
electron-positron beam.
The dynamical properties of each constituent have been thoroughly
analyzed in previous works
(e.g. \citealt{sha73,hen91,fer95,mar97,ren98,sau03,sau04}), but it is the first
time where they are invoked altogether as necessary ingredients to
reproduce the different spectral states of a same object.  

We showed that the various canonical states can be qualitatively
explained by varying {\em independently} the transition radius $r_J$ and
the disc accretion rate $\dot m$. In our view, the Quiescent and Hard
states are dominated by non relativistic jet production from the JED,
providing henceforth a persistent synchrotron jet emission. The Soft
state is obtained when the transition radius $r_J$ becomes smaller than
the last marginally stable orbit $r_i$, a SAD is established throughout
the whole accretion disc. Intermediate states, between Hard and Soft, are
expected to display quite intricate and variable spectral energy
distributions. Luminous Intermediate states, obtained during the
Hard-to-Soft transitions, are those providing the unique conditions for
intermittent pair creation. These pairs give rise to a ultra relativistic
beam propagating on the MHD jet axis, explaining both the observed
superluminal motions and hard energy tail. \\

The dynamical structure presented here (JED, SAD, MHD jet and,
occasionally, a pair beam) seems to be consistent with all available
information about the canonical spectral states of BH XrBs. However, a
more quantitative analysis is critical. In particular, we need to show
that the base of the MHD jet can indeed provide a hot corona with the
correct spectral signature. Then, a precise estimate of the radio/X-ray
correlation predicted by our model and its comparison to observations
(through the computation of SEDs) will be a test of prime importance for
its validity (Petrucci et al., in preparation).\\

In our view, the magnetic flux available at the inner disc regions is a
fundamental and unavoidable ingredient that most probably varies from one
system to another. We believed that it controls, with $\mdot$, the
evolution of the transition radius $r_J$, through the variation of the
magnetization $\mu$ in the different part of the accretion disc. The
variation of $\mu$ and $\mdot$ all along the HID diagram would explain
the variation of $r_J$. This is a work in progress (Ferreira et al. in
preparation).\\

{\it Acknowledgment: We would like to thanks Marek Gierlinski for the
  wonderful organization of this very interesting workshop.}




\end{document}